# The role of fabrication deviations on the photonic band gap of 3D inverse woodpile nanostructures

Léon A. Woldering,[1,*] Allard P. Mosk,[1] R. Willem Tjerkstra,[1] and Willem L. Vos[1,2]

[1]*Complex Photonic Systems (COPS),*

*MESA+ Institute for Nanotechnology,*

*University of Twente, P.O. Box 217,*

*7500 AE Enschede, The Netherlands*

[2]*Center for Nanophotonics, FOM Institute for Atomic and Molecular Physics (AMOLF),*

*Kruislaan 407, 1098 SJ Amsterdam, The Netherlands*

(Dated: December 24, 2008, version 1)

## Abstract

In this report the effects of unintended deviations from ideal inverse woodpile photonic crystals on the band gap are discussed. These deviations occur during the nanofabrication of the crystal. By computational analyses it is shown that the band gap of this type of crystal is robust to most types of deviations that relate to the radii, position and angular alignment of the pores. However, the photonic band gap is very sensitive to tapering of the pores, *i.e.*, conically shaped pores instead of cylindrical pores. To obtain three-dimensional inverse woodpile photonic crystals with a large volume, our work shows that with modern fabrication performances, tapering contributes most significantly to a reduction in the photonic strength of inverse woodpile photonic crystals.

## I. INTRODUCTION

Three-dimensional photonic band gap crystals can be considered as supreme devices to control light. It is predicted that spontaneous emission of light is inhibited in these structures[1] or that light is localized.[2] Another intriguing aspect of photonic crystals is that an embedded point defect acts as a microcavity wherein photons can be stored or slowed down.[3] Furthermore, photonic crystal waveguides can be used to "mold the flow of light".[4]

A photonic crystal is a periodic structure made from two different materials that are periodically alternated over length scales in the order of the wavelength of light. Typically one material is air and the other a semiconductor with a high index of refraction. Light with certain wavelengths and directions cannot propagate in such a structure because of Bragg diffraction.[5] In three-dimensional photonic band gap crystals, certain frequencies of light are forbidden from propagating in any direction. For this to occur, the photonic strength of the crystal must be high enough.[6]

Different types of three-dimensional photonic crystals have been conceived,[7,8] but of particular interest are those that potentially provide large three dimensional photonic band gaps. Such structures offer ultimate control of light in all three dimensions simultaneously and therefore many research groups are studying their fabrication.[9–14]

A promising class of three-dimensional photonic crystals have a diamond-like symmetry, in particular with an inverse woodpile structure.[15] Inverse woodpile photonic crystals are very interesting because of their conceptual ease of fabrication and a broad photonic band gap that is robust to disorder and fabrication imperfections. These crystals consist of two perpendicular arrays of cylindrical pores of air in a high refractive index material, see Figure 1(A). In this paper we consider silicon as high refractive index material since it is conveniently available with high purity. Advanced nanofabrication methods exist for silicon, and it is used in electronic circuits which promises integration.

The calculated band gap of inverse woodpile photonic crystals has a relative band width of more than $(\Delta\omega/\omega) = 25\%$.[15,16] In practice the applied fabrication methods impose restrictions on the quality of the crystals. To elaborate, all fabrication techniques induce undesired deviations from the intended ideal crystal structure. The occurrence of these unintended deviations during fabrication is unavoidable. These deviations influence the interaction of light with the structures and typically reduce the width of the band gap, as was shown, for example, for a woodpile photonic crystal.[17] Therefore, it is important to know what the expected effect is on the photonic band gap of these deviations from an ideal inverse woodpile crystal geometry. In this report the effect of several types of deviations is discussed that most frequently occur as a result of the manufacturing process. These deviations are investigated by calculations. The obtained knowledge aids process selection and process optimization such that fabricating the desired inverse woodpile structures can be performed more efficiently.

## II. IDEAL STRUCTURE

Figure 1 shows an "ideal" cubic inverse woodpile crystal. This crystal consists of two perpendicular, geometrically identical, sets of cylindrical pores, as first proposed by Ho $et\ al.$,[15] see Figure 1(A). The important crystal dimensions pore radius $R$, and lattice parameters $a$ and $c$ are indicated in Figure 1(B). The individual pore sets have a centered rectangular lattice symmetry,[18] with $(a/c) = \sqrt{2}$. The combination of the two sets results in a three-dimensional structure with a face-centered-cubic lattice. The centers of one perpendicular pore set are exactly aligned between rows of pores of the other set. The pores of the two sets overlap, as is clearly visible in Figure 1(A). When the ratio of $(R/a)$ is equal to 0.24, a crystal has a photonic band gap with a maximum relative width of more than $(\Delta\omega/\omega) = 25\%$.[15,16] Furthermore the pores should be perfectly cylindrical and not tapered, *i.e.*, not conical. The pore walls should be smooth to avoid any unwanted scattering of light.[19]

We calculated photonic band structures of such an ideal crystal with the MIT photonic bands package.[20] This program calculates eigenstates and eigenvalues of Maxwell's equations in periodic, infinite structures. The program uses a planewave basis. For these calculations the structure is considered to have a orthorhombic lattice symmetry. The band structures were calculated with a grid-resolution of $\frac{c}{\Delta x} \times \frac{a}{\Delta y} \times \frac{c}{\Delta z} = 70 \times 100 \times 70$ and 73 k-points. In the appendix we show that such a grid-resolution offers enough accuracy for our calculations. The dielectric constant of silicon that we used is $\varepsilon_{Si} = 12.1$.

In Figure 2(A) calculated bands 1 to 8 are shown. Figure 2(B) shows the Brillouin zone of the orthorhombic lattice. For $(R/a) = 0.24$ and $(a/c) = \sqrt{2}$, our calculations confirmed a relative width of the band gap of $(\Delta\omega/\omega) = 25.3\%$, centered around frequency $= 0.564\,[\omega a/2\pi c]$ (here $c$ is the speed of light), which is equal to results by Hillebrand *et al.*[16]

### III. OVERVIEW OF DEVIATIONS

Various techniques can be used to etch the two separate pore sets 1 and 2. Examples are photo electrochemical etching,[10,21,22] reactive ion etching,[12,23] and focused ion beam milling.[9,14] Fabrication of the second set of pores is challenging, as it is difficult obtain two sets of pores with equal radii and to correctly align the second set exactly between the first set of pores. Furthermore, it is necessary that the pores are perpendicular to each other, which imposes further challenges to the alignment of the second pore set to the first. Finally, it is important to maintain the homogeneity of the photonic crystal throughout its entire volume. Therefore the fabricated pores must be cylindrical.

In this paper several types of deviations are discussed that occur when a nanostructure is made from 2 sets of pores:

a) The ratio radius $R$ to lattice parameter $a$ ratio is different from 0.24.

b) The two sets of pores have different cylinder radii.

c) The two sets of pores are slightly displaced.

d) The pores are not perpendicularly aligned with respect to each other.

e) Tapered, *i.e.*, conical pores versus the desired cylindrical ones.

Furthermore we calculate the relative width of the band gap resulting from combinations of *1)* the ratio of radius $R$ to lattice parameter $a$ and *2)* the ratio of lattice parameters $a$ to $c$.

## IV. RESULTS AND DISCUSSION

### A. Radius $R$ to lattice parameter $a$ different from 0.24

When the radii of the etched pores deviate from the intended radius, the ratio of radius $R$ to lattice parameter $a$ is different from the optimal value of $(R/a) = 0.24$. This is discussed by Ho *et al.* and Hillebrand *et al.*[15,16] They showed that already at $(R/a)$ ratios of around 0.18 and 0.28 the relative width of the band gap has been reduced by half. At $(R/a) < 0.14$ and $(R/a) > 0.29$, there is no longer a band gap present. In order to obtain data with known increments of $(R/a)$ *and* to check of our computation method, we calculated the band structures for the region $0.14 < (R/a) < 0.30$; see Figure 3.

The maximum relative width of the band gap of $(\Delta\omega/\omega) = 25.4\%$ is found for $(R/a) = 0.245$. Here the central frequency of the band gap equals 0.581 $[\omega a/2\pi c]$. For $(R/a)$ values away from the optimal value, the relative band gap width is reduced. The band gap moves to higher central frequencies with increasing $(R/a)$. In comparison to the values reported in literature[16] we find good agreement to within $(\Delta\omega/\omega) = \pm 1.8$ percent point in the entire calculated range (not shown in Figure 3). Thus, we conclude that our calculations are accurate.

The ratio $(R/a)$ directly relates to the volume fraction of silicon $\phi_{Si}$, *i.e.*, increasing $(R/a)$ results in lower $\phi_{Si}$. At $(R/a) = 0.245$, the volume fraction of silicon $\phi_{Si} = 19.8\%$, which means that the amount of high refractive index material is lower than the amount of low refractive index material. This is consistent with the heuristic rule that *"The optical path lengths in the high and low index materials should be equal, to enhance interference"* that generally applies to photonic crystals.

### B. A $\Delta$y displacement between the two sets of pores

One of the alignment errors that may occur during fabrication is that the pores in the second set are not positioned exactly in between the pores of the first pore set. This results in a $\Delta$y displacement as illustrated in Figure 4, where the second set is misaligned to the left or the right from its ideal position. The first set of pores is pointing "into" the paper and the second set of pores runs parallel to the z-axis. Due to symmetry reasons, a misalignment to the left will have the same effect on the photonic band gap as a misalignment to the right. Such a $\Delta$y displacement typically results from the lithographic process by which the second pore set is defined. Alternatively, it results from a misalignment during focused ion beam milling of the structures. It is impossible to obtain perfect alignment, and a small deviation from the ideal geometry is inevitable. In literature the effect of this deviation on the band gap is discussed.[24,25] A displacement of up to around 10% of the lattice constant results in a slight reduction of the relative photonic band width to > 90% of the original value. This means that the relative width of the band gap that remains is more than $(\Delta\omega/\omega) = 22.8\%$. When the $\Delta y$ displacement exceeds 10% of the lattice constant, the relative width of the band gap decreases rapidly. Since contemporary alignment processes perform much better than this required precision, we do not expect this deviation to contribute significantly to a reduced relative width of the band gap.

### C. Different cylinder radii of the two sets of pores

When two separate processes are used to obtain the two perpendicular pore sets, it is very challenging to obtain two sets of pores with equal cylinder radii. Many scenarios can be distinguished how the radii of the two pore sets differ, two of which will be discussed here:

a) the radius of the first pore set is according to specification ( $(R_1/a) = 0.24$ ), but the radius of the second pore set is varied ( $(R_2/a) \neq 0.24$ ).

b) The first and second pore set both differ from ideal $(R/a) = 0.24$, with $(R_1/a)$ larger and $(R_2/a)$ equally smaller than specified for an ideal structure.

With the MIT photonic bands package we have calculated the effects of these deviations on the relative width and central frequency of the band gap. By analyzing the calculated data, whilst choosing as a criterion that a photonic band gap width of $(\Delta\omega/\omega) > 20\%$ is acceptable, we can quantify the robustness of the crystal structure towards these geometrical deviations. In both scenarios lattice parameters $a$ and $c$ are constant.

a) Figure 5 shows the the band gap edges and the relative width of the band gap versus $\Delta(R/a)$. Difference $\Delta(R/a)$ is defined as:

$$\Delta\left(\frac{R}{a}\right) = \left(\frac{R_2}{a}\right) - \left(\frac{R_1}{a}\right). \tag{1}$$

The central frequency of the band gap is lowest at small $\Delta(R/a)$, and increases with increasing $\Delta(R/a)$. A maximum relative width of the band gap is found for $\Delta(R/a) = 0$. The relative width of the band gap decreases as $\Delta(R/a)$ differs from zero. The band gap is most sensitive to an increasing ratio $(R_2/a)$. When $\Delta(R/a) = +0.048$, the relative width of the band gap decreases to less than 5%. In contrast, when $\Delta(R/a) = -0.048$, the relative band gap width remains 15%. We conclude that our criterion of a minimal relative width of the band gap of $(\Delta\omega/\omega) = 20\%$ is found for $\Delta(R/a) \leq \pm 0.024$. This corresponds to a large difference in the etched radius $\Delta R \leq \pm 16$ nm. Typically the achieved radius will be closer to the target radius.

b) Figure 6 shows the the band gap edges and the relative width of the band gap versus $\Delta(R_{(1+2)}/a)$. The difference in ratios $\Delta(R_{(1+2)}/a)$ is defined as:

$$\Delta\left(\frac{R_{(1+2)}}{a}\right) = \left(\frac{R_1}{a}\right) - 0.24 \; and \; \Delta\left(\frac{R_{(1+2)}}{a}\right) = 0.24 - \left(\frac{R_2}{a}\right). \tag{2}$$

The relative width of the band gap decreases as $\Delta(R_{(1+2)}/a)$ differs from zero. When the pore radii are changed by $\Delta(R_{(1+2)}/a) = +0.035$, the relative width of the band gap is halved. For $\Delta(R_{(1+2)}/a) = +0.05$, the band gap vanishes. We conclude that our criterion of a minimal relative width of the band gap of $(\Delta\omega/\omega) = 20\%$ is found for $\Delta(R_{(1+2)}/a) \leq 0.018$, which corresponds to large difference in the etched radius of around $R \leq 12$ nm. However, the achieved radius will typically be much closer to the target radius. In addition, Figure 6 shows that the central frequency of the band gap remains at almost the same position as $\Delta(R_{(1+2)}/a)$ increases, which is convenient: even when this deviation occurs during fabri-cation of the crystal, it will nonetheless be active at the optical frequency it was designed for.

From a) and b) we conclude that inverse woodpile photonic crystals are robust to differences between optimized $(R_1/a)$ and $(R_2/a)$ that may occur as a result of the fabrication process. A qualitative result that is reported in literature agrees with our extensive calculations.[16] Furthermore, in modern fabrication the margins of the deviations are expected to remain well within these limits.

### D. Angular misalignments of the two pore directions

The second set of pores that is going to be fabricated may be angularly misaligned with the respect to the first pore set that is already present. This means that the first pore set is not perpendicular to the etching direction of the second pore set. For example, in experiments where the second pore direction is milled using a focused ion beam setup, misalignment depends on how well the sample is positioned with respect to the ion beam. Figure 7 shows the three possible directions in which the two pore sets can be misaligned. The x-axis is the ideal direction of the first pore set and the z-axis the ideal direction of the second pore set. The following three angular misalignments are discriminated:

a) Misalignment by angle $\alpha$ occurs when the second set of pores, which is perpendicular to the first set of pores, is etched when the first set of pores is tilted about the y-axis out of the plane perpendicular to the etching direction.

b) Misalignment by angle $\beta$ occurs when the pattern of the second set of pores is rotated about the z-axis with respect to its correct positioning.

c) Misalignment by angle $\gamma$ occurs when the second set of pores is etched perpendicular to the first set, but the pattern of the second set is rotated about the first pore axis.

Misalignment by angles $\gamma$ and $\beta$ break the long-range periodic order of the three-dimensional structure. For this reason we can not determine the effect of these misalignments on the photonic band gap with the MIT photonic bands package, as band structures can only describe periodic structures.[26] Nonetheless we can illustrate to what extent these misalignments break the long-range periodic order of the structure by calculating after how many lattice spacings $c$ along the x- or z-direction, the structure deviates by one lattice spacing $a$, see Figure 8. With increasing misalignment angle, the pores need less lattice spacings $c$ to traverse one lattice spacing $a$, which indicates increasing non-crystallinity of the structure. For an angle $\beta$ or $\gamma < 0.5°$, the structure is periodic over $> 180$ lattice spacings $c$. In fabrication practice where the lattice spacing $a$ is set at 680 nm and the obtainable pore depths are around $h = 10$ $\mu$m, the calculated periodicity for misalignment angles $\gamma$ and $\beta < 0.5°$ can still be considered long-range. Since a better method to quantify the effect of this misalignment lacks, the upper limit for these two misalignments individually must be determined depending on the depth of the pores that can be obtained and the chosen lattice parameters.

The situation where $\beta = \gamma = 0°$ and $\alpha \neq 0°$ is discussed in detail in literature.[24,25] Increasing angle $\alpha$ results in a lower width of the band gap. The effect of an increasing misalignment that can be afforded is limited to an angle of about 5°: a change from $(\Delta\omega/\omega) = 24\%$ at $\alpha = 0°$ to $(\Delta\omega/\omega) = 21\%$ at $\alpha = 5°$ is reported in this reference, with a slightly increasing center frequency of the band gap. This result is similar to results described for a crystal of high refractive index woodpiles.[17] Since $\alpha$ is typically expected to remain far below 5°, the effect of this misalignment can be neglected.

### E. Conical pores

During the fabrication of pores for a three-dimensional crystal by means of, *e.g.*, focused ion beam milling and reactive ion etching, it is common for the pores to be conically shaped. This so-called tapering has a pronounced effect on the band gap of the crystal. Since we did not find a detailed analysis of the effect of this deviation in literature, it will be further discussed in this section. Three situations may be distinguished:

a) the crystal has one set of pores which displays tapering, whereas the other set is cylindrical,

b) both pore sets display the same amount of tapering, and

c) both pore sets are tapered, but with a different amount.

Here we will consider situation b) as it provides a general representation of the deviation and is the most straightforward case to analyze. Figure 9 shows a structure consisting of 10 pores by 10 pores with tapered pores in both directions. In the third dimension the structure is assumed to be (infinitely) large, therefore we limit our discussion to the x-z plane. In the center of the inverse woodpile (black dot, [x, z] = [0, 0]) the radii of both pores are equal to the desired radius. On any other position the ratio $(R/a)$ of the two pore sets is different. Along the diagonal from bottom left to top right the radii of the two pore sets are equal, but different at all positions. This changes one quantity of the structure: the ratio $(R/a)$ decreases, hence the volume fraction of silicon increases and thus the effective refractive index $n_{eff}$ is increased. Consequently, the width and the central frequency of the band gap are changing along this diagonal: the relative width of the band gap has an optimum at [0, 0], and the central frequency is chirped and decreases when going from bottom left to top right. Unfortunately it is strictly speaking not possible to calculate band structures of tapered inverse woodpiles. However, with the following theoretical analysis, the effect of the tapering on the gap can be estimated.

The model photonic structure under consideration has an ideal geometry at its center with $(R/a) = 0.24$. There, it has a band gap at frequency $\omega_{gap} = 0.564\ [\omega a/2\pi c]$. For non-tapered crystals with equal radii of both pore sets, we have calculated in section IV A how the relative width of the band gap depends on $(R/a)$, keeping the lattice parameters $a$ and $c$ fixed. To estimate the effect of tapering on the band gap, the data in Figure 3 were used to calculate the upper and lower band edges of the "local photonic band gap" along the diagonal in structures which have different amounts of tapering.[27] We calculated from -23 to 23 unit cells along the diagonal of the structure, *i.e.*, from [-23, -23] to [23, 23] in Figure 9 (all coordinates are in units [# of unit cells, # of unit cells] ).

Figure 10 shows the results of these calculations for four different amounts of tapering, namely 1°, 0.5°, 0.25°, and 0.125°. Each plot shows the upper and lower band edges versus position along the diagonal in the crystal. The horizontal dashed lines are at frequency $\omega_{gap} = 0.564\ [\omega a/2\pi c]$, for which these inverse woodpiles are optimized. The Figure shows that the photonic structures are extremely sensitive to tapering of both pore sets. For a tapering as little as 1° the relative gap width rapidly decreases when moving away from the center, see Figure 10(A). With decreasing tapering, the relative width of the band gap decreases less rapidly when moving away from [0, 0], see Figures 10(B) to (D).

As a measure of quality, the transmittance $T$ along the diagonal is calculated for light at the frequency of the band gap center. A low transmittance is indicative of a strongly photonic behaviour and *vice versa*. Transmission $T$ depends on the Bragg attenuation length $L_B$ and can be calculated using equation 24 from the work by Spry and Kosan[28]:

$$T = \left[\cosh\left(\frac{h_{bands}}{L_B}\right)\right]^{-2}. \qquad (3)$$

In structures without tapering $h_{bands}$ is equal to the thickness of the photonic crystal. In case of tapering, $h_{bands}$ is the length along the diagonal for which a band gap for the optimal frequency exists. In other words, as $h_{bands}$ increases, the volume of the inverse woodpile with a band gap increases. In Figure 10 $h_{bands}$ is determined by the points were the band gap edges cross the horizontal dashed line displaying the relative frequency for which these structures are optimized.

The Bragg attenuation length $L_B$ is a measure for the distance that light can penetrate into the photonic crystal before being reflected by the lattice planes. This length is derived from the photonic strength $S^{29}$:

$$L_B = \frac{\lambda_{Bragg}}{\pi S}. \tag{4}$$

It is obvious that a short Bragg attenuation length pertains to crystals with a high photonic strength $S$. In order to estimate an "effective" Bragg length $L_B$ for all directions in the ideal inverse woodpiles, the photonic strength $S$ in all directions is chosen to be equal to the relative width of the photonic band gap, therefore $(\Delta\omega_{gap}/\omega_{gap}) = S = 25.3\%$. Finally we take $\lambda_{Bragg}$ equal to the central wavelength of the calculated band gap $\lambda_{gap}$ for the ideal geometry. The calculated Bragg attenuation length equals $L_B = 2.2$ unit cells. Therefore, in the center of each plot in Figure 10 a horizontal bar equal to the calculated Bragg attenuation length $L_B$ is shown.

For a structure with $1°$ tapering, $h_{bands} = 3.1$ unit cells, see Figure 10(A). The calculated transmittance is as high as $T = 0.22$, which indicates that the structure has little effect on the transmission of light. This structure can hardly be considered a photonic band gap crystal. For $0.5°$ tapering $h_{bands}$ is limited to $h_{bands} = 6.2$ unit cells, see Figure 10(B). Therefore only a small volume of the inverse woodpile has a photonic band gap. This results in a transmittance of $T = 0.015$, which is 15 times better compared to $1°$ tapering. Nonetheless, from the values calculated for $h_{bands}$ it is obvious that when the fabrication process yields pores with a tapering of 0.5 to 1°, it is not very useful to try to obtain pores deeper than a few micrometer, as pores deeper than $h_{bands}/\sqrt{2}$ do not contribute to an effect of the photonic band gap. In the best case reported in Figure 10, we see that at a tapering of 0.125° the transmittance is extremely low at $T = 1.2 \cdot 10^{-9}$, which shows that the band gap has a significant effect on the transmittance. In addition, $h_{bands} = 24.5$ unit cells. This means that a photonic band gap exists in a 10 x 10 $\mu$m$^2$ inverse woodpile crystal with a lattice spacing $a$ equal to around 680 nm. Therefore, trying to obtain pores deeper than a few micrometer is useful, as deeper pores contribute to stronger photonic crystals until a depth of 10 $\mu$m is reached.

On all positions in the model inverse woodpile that are not on the diagonal from bottom left to top right, the radii of the two pore sets are different. Unfortunately calculating the "local photonic band gap" properties for all these positions would be extremely time-consuming, which limits further exploration of the problem *and* limits an analysis of situations a) one pore set with tapering, and c) two pore sets with different amounts of tapering. However, the above analysis of the local gap along the [-23, - 23] to [23, 23] diagonal of the tapered inverse woodpile, which is the worst-case in that structure, clearly illustrates the importance of reducing the tapering as much as possible.

In our experiments to date, we have been able to produce cylindrical pores with tapering values as low as 0.20°, see Figure 11. This set of pores is oriented in a centered rectangular lattice. They were etched using an optimized reactive ion etch based on our earlier work.[23] These pores have a depth of more than $h = 7$ $\mu$m with a lattice spacing of 500 nm. The pore diameter is $D_{pore} = 364$ nm, which means that these pores have a high aspect ratio of more than $A = 20$. A three-dimensional inverse woodpile photonic crystal made by consecutive etching of two such sets of pores would span 14 unit cells along the x and z-axis. Figure 10 clearly shows that structures fabricated with pores of these dimensions are very interesting to consider. Such a fabricated inverse woodpile is expected to be strongly photonic and may even display a band gap. Nevertheless, it is also evident that to obtain an increased relative width of the band gap or to make larger photonic crystals with even deeper pores the tapering needs to be reduced further.

### F. Optimization by using deviations

In literature an inverted woodpile photonic crystal is described, fabricated from ellipsoidal pores.[13] There, a larger band gap is predicted when the ratio $(a/c)$ in their structure is increased from $\sqrt{2}$ to 1.63. Motivated by this result, we have investigated whether the ratio $(a/c) = \sqrt{2}$ is optimal for inverse woodpile photonic crystals.

Using the MIT photonic bands package we calculated the relative width of the band gap for structures with varying ratios of $(a/c)$. Since the optimal $(R/a)$ also varies with the chosen $(a/c)$, we also varied $(R/a)$ in our calculations. The ratio $(a/c)$ was varied from 1 to 1.8, with increments as small as $|\Delta(a/c)| = 0.025$ in the region of the maximum relative width of the band gap. The ratio $(R/a)$ was varied from 0.18 to 0.28. with increments of $|\Delta(R/a)| = 0.0025$ in the region of the maximum relative width of the band gap.

Figure 12 shows an interpolated three-dimensional surface- and contour plot of the calculated band gaps versus the ratios $(R/a)$ and $(a/c)$. The plot was interpolated using the Renka-Cline gridding method in Origin 7. Figure 12 shows that there is a large region of $(a/c)$ and $(R/a)$ combinations where the relative band gap width is much larger than $(\Delta\omega/\omega) = 26.5\%$, see the black circular area around $(a/c) = 1.6$ and $(R/a) = 0.23$. Naively we expected an optimal result for the case where two hexagonal patterns of pores are combined ( $(a/c) = \sqrt{3}$ ), due to the high symmetry of the hexagonal lattice. However, Figure 12 clearly shows that this is not the case.

The largest band gap is found for $(a/c) = 1.575$ and $(R/a) = 0.23$. At these values, a relative width of the band gap of $(\Delta\omega/\omega) = 26.9$ % is calculated, slightly larger than the band gap of $(\Delta\omega/\omega) = 25.3$ % for the structure with $(a/c) = \sqrt{2}$ and $(R/a) = 0.24$ that we hitherto referred to as ideal (cubic) crystal. Our result reproduces results reported in literature.[30] The increase in relative band gap width is more than $\Delta(\Delta\omega/\omega) = 1.6$ %pt, and can be obtained by straightforward modification of the lithographic processes in the fabrication route of inverse woodpile photonic crystals.

Figure 13 shows the calculated band structures of the improved structure with $(a/c) = 1.575$ and $(R/a) = 0.23$, and the calculated band structures of the crystal with $(a/c) = \sqrt{2}$ and $(R/a) = 0.24$. For the optimal inverse woodpile photonic crystal, the frequencies of the band edges are higher. The central frequency of this structure is $\omega = 0.635$ $[\omega a/2\pi c]$, compared to $\omega = 0.564$ $[\omega a/2\pi c]$ for the cubic structure. This difference is due to the decreased volume fraction of silicon in the structure which results in a lower effective refractive index $n_{eff}$. When comparing the band structures of both types of crystal, we observe that the lower 4 bands are scaled along the frequency axis because of the different effective refractive indices. For most wave-vectors in the Brillouin zone, the upper bands are also scaled, with the exception of the $\Gamma$ - $Y$, $\Gamma$ - $X$, $\Gamma$ - $U$ and $\Gamma$ - $Z$ directions, which are clearly different. Apparently these are the directions that are most influenced by the changing geometry of the structure.

## V. CONCLUSIONS

In this paper we have discussed possible unintended deviations from an ideal inverse woodpile photonic crystal geometry that may occur in the fabrication process of such struc-tures. We have shown that the photonic band gap of this type of crystals is robust to most types of fabrication deviations that relate to the radii, position and angular alignment of the pores, within the tolerances of modern fabrication processes. Therefore, with regards to these fabrication deviations, high quality crystals are expected. We do find that these crystals are very sensitive to tapering of the pores, *i.e.*, conically shaped pores instead of cylindrical.

Our analysis shows that for tapering values that are obtainable in practice with reactive ion etching, these structures are still expected to display a band gap. To obtain three-dimensional inverse woodpile photonic crystals with a large volume, our work shows that tapering is arguably the most important deviation to minimize. In fact, tapering of the pores needs to be reduced to values as low as possible.

By closely looking at the lattice parameters of the inverse woodpile photonic crystal made from perfectly cylindrical pores, we calculated that an alternative geometry potentially provides a slightly larger photonic band gap. It is straightforward to modify lithographic processes to pursue the fabrication of this geometry.

## VI. APPENDIX: GRID RESOLUTION

One of the most important parameters that can be set in the MIT photonic bands program is the grid-resolution with which the unit cell is defined. Preferably this is set to high numbers to have the highest accuracy possible. However, setting the grid-resolution to higher values means that the program will use more memory and take longer to finish the calculation. Therefore, in order to find an optimal setting it is worthwhile to investigate the output of the calculations versus grid-resolution. To the best of our knowledge such an extensive investigation has not been reported before.

TABLE I: Overview of grid-resolutions set in different calculations for an inverse woodpile with $(R/a) = 0.24$ and $(c/a) = \sqrt{2}$. For grid-resolution 8, marked with an asterisk, the MIT photonic bands program only calculated the volume fraction. Further calculations of the frequency limits of the band gap were not possible due to memory-constraints.

| Calculation [#] | Grid resolution [$\frac{c}{\Delta x} \times \frac{a}{\Delta y} \times \frac{c}{\Delta z}$] |
|---|---|
| 1 | 12×17×12 |
| 2 | 17×24×17 |
| 3 | 24×34×24 |
| 4 | 34×48×34 |
| 5 | 48×68×48 |
| 6 | 68×96×68 |
| 7 | 96×136×96 |
| 8* | 136×192×136 |

We performed several calculations for an inverse woodpile with a ratio of $(R/a) = 0.24$ and a ratio of $(c/a) = \sqrt{2}$. The grid-resolution was set to the values reported in Table I. We show how the volume fraction and the upper- and lower frequency limits of the band gap converge with increasing grid-resolution.

We compared volume fraction of silicon calculated with the MIT photonic bands program to the exact volume fraction of the structure that was calculated analytically. Figure 14 shows how the calculated volume fraction depends on the grid-resolution. The horizontal dashed-dotted line displays the exact volume fraction of silicon in the inverse woodpile. At low grid-resolutions the calculated volume fraction is too low, which indicates that the unit cell is poorly defined. Increasing the grid-resolution increases the volume fraction, as is expected. After an "overshoot", the calculated volume fraction of silicon converges to its exact value. We conclude that at a grid-resolution of $\frac{c}{\Delta x} \times \frac{a}{\Delta y} \times \frac{c}{\Delta z} = 68 \times 96 \times 68$ and onward the unit cell is defined with high-enough accuracy.

Next we study how the calculated band gap depends on the grid resolution. In Figure 15 is shown that at lower grid-resolutions the band gap limits are elevated. The frequency limits reduce to lower values at higher grid-resolutions. Increasing from a grid-resolution of $\frac{c}{\Delta x} \times \frac{a}{\Delta y} \times \frac{c}{\Delta z} = 48 \times 68 \times 48$, the frequency limits of the band gap are constant within 0.5%. Combined with the data for the volume fraction of silicon we conclude that the grid-resolution of $\frac{c}{\Delta x} \times \frac{a}{\Delta y} \times \frac{c}{\Delta z} = 70 \times 100 \times 70$ that we used throughout this report is sufficiently high for our purposes.

**Acknowledgements**

We thank Philip Harding for fruitful discussions and Bas Benschop for assistance. This research was supported by NanoNed, a nanotechnology program of the Dutch Ministry of Economic Affairs, and this work is part of the research program of the Stichting voor Fundamenteel Onderzoek der Materie (FOM), which is financially supported by the Nederlandse Organisatie voor Wetenschappelijk Onderzoek (NWO). This work is also supported by a VICI fellowship from the Nederlandse Organisatie voor Wetenschappelijk Onderzoek (NWO) to WLV.

[*] Electronic address: `l.a.woldering@utwente.nl`

[1] E. Yablonovitch, Phys. Rev. Lett. **58**, 2059 (1987).

[2] S. John, Phys. Rev. Lett. **58**, 2486 (1987).

[3] K. J. Vahala, Nature **424**, 839 (2003).

[4] J. D. Joannopoulos, R. D. Meade, and J. N. Winn, *Photonic crystals - molding the flow of light*, Princeton University Press, Princeton, New Jersey, 1995.

[5] W. L. Bragg, Proc. Camb. Phil. Soc. **17**, 43 (1913).

[6] The photonic strength $S$ is a measure of the interaction strength of photonic crystals with light. A higher $S$ means that the crystal is more strongly interacting with light.

[7] C. López, Adv. Mater. **15**, 1679 (2003).

[8] M. Maldovan and E. L. Thomas, Nature Mat. **3**, 593 (2004).

[9] K. Wang, A. Chelnokov, S. Rowson, P. Garoche, and J. M. Lourtioz, J. Phys. D: Appl. Phys. **33**, L119 (2000).

[10] J. Schilling et al., J. Opt. A: Pure Appl. Opt. **3**, S121 (2001).

[11] J. Schilling et al., Appl. Phys. Lett. **86**, 011101 (2005).

[12] S. Takahashi, M. Okano, M. Imada, and S. Noda, Appl. Phys. Lett. **89**, 1231061 (2006).

[13] M. Hermatschweiler, A. Ledermann, G. A. Ozin, M. Wegener, and G. von Freymann, Adv. Funct. Mater. **17**, 2273 (2007).

[14] R. W. Tjerkstra, F. B. Segerink, J. J. Kelly, and W. L. Vos, J. Vac. Sci. Technol. B **26**, 973 (2008).

[15] K. M. Ho, C. T. Chan, C. M. Soukoulis, R. Biswas, and M. Sigalas, Solid State Comm. **89**, 413

(1994).

[16] R. Hillebrand, S. Senz, W. Hergert, and U. Gösele, J. Appl. Phys. **94**, 2758 (2003).

[17] A. Chutinan and S. Noda, J. Opt. Soc. Am. B **16**, 240 (1999).

[18] C. Kittel, *Introduction to solid state physics*, Wiley, London, 1967.

[19] A. F. Koenderink, A. Lagendijk, and W. L. Vos, Phys. Rev. B. **72**, 153102 (2005).

[20] S. G. Johnson and J. D. Joannopoulos, Opt. Express **8**, 173 (2001).

[21] S. Rowson, A. Chelnokov, and J. M. Lourtioz, Electronics Letters **35**, 753 (1999).

[22] S. Matthias, F. Müller, C. Jamois, R. B. Wehrspohn, and U. Gösele, Adv. Mater. **16**, 2166 (2004).

[23] L. A. Woldering, R. W. Tjerkstra, H. V. Jansen, I. D. Setija, and W. L. Vos, Nanotechnology **19**, 145304 (2008).

[24] J. Schilling and A. Scherer, Photonics and Nanostructures **3**, 90 (2005).

[25] In reference[24] an inverse woodpile is considered with slightly different geometrical parameters.

[26] A calculation is possible by using supercell input. However, the supercells that are needed for these calculations would be too large to be practically possible to use due to memory restrictions.

[27] We used $9^{th}$ order polynomal interpolation to calculate the properties along the diagonal. The polynomal constants were determined with 8 digit precision.

[28] R. J. Spry and D. J. Kosan, Appl. Spectroscopy **40**, 782 (1986).

[29] A. F. Koenderink, *Emission and transport of light in photonic crystals*, PhD thesis, University of Amsterdam, 2003.

[30] R. Hillebrand and W. Hergert, Photonics and Nanostructures **2**, 33 (2004).

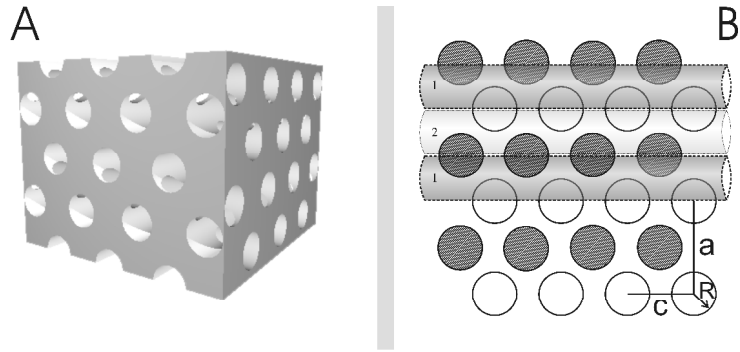

FIG. 1: A) Three-dimensional representation of a cubic inverse woodpile photonic crystal. The alignment of the pores between the pores of the other set can be clearly seen, as well as the overlap of the pores. B) Face-on view of the pores in the inverse woodpile photonic crystal. The centers of the perpendicular cylinders are aligned exactly between rows of cylinders of the other set. The individual pore sets have a centered rectangular lattice symmetry, with lattice parameters $(a/c) = \sqrt{2}$. An optimal photonic crystal is achieved when the ratio of radius $R$ to lattice parameter $a$ equals 0.24.

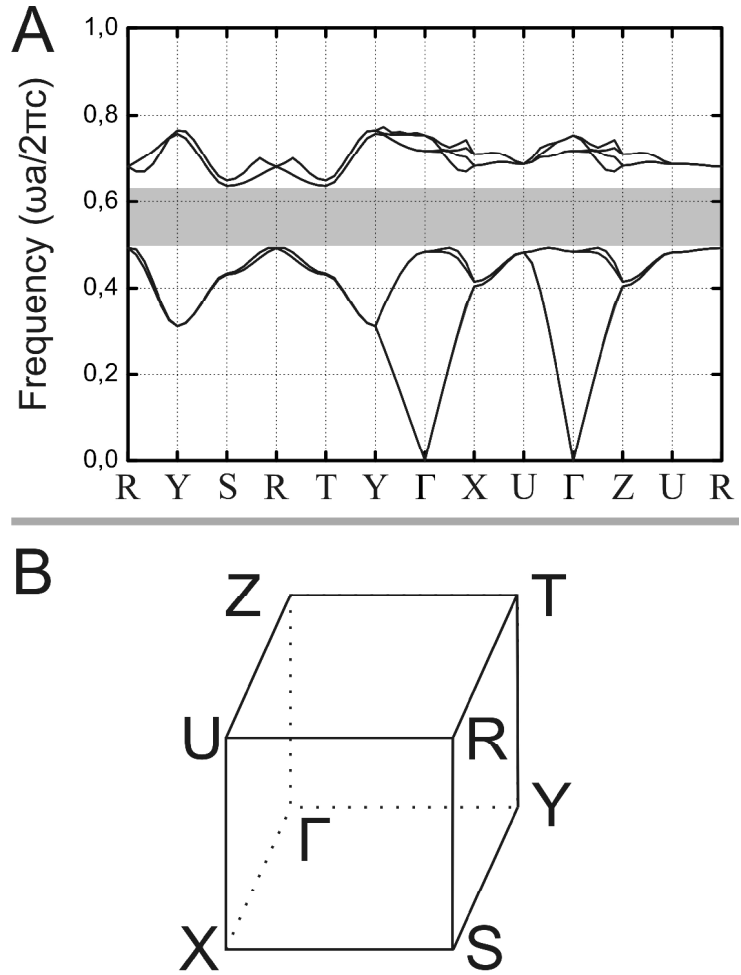

FIG. 2: A) Calculated band structure of an ideal inverse woodpile photonic crystal, with the radii $R$ of the two perpendicular pore sets equal to 0.24 times the lattice parameter $a$, and $(a/c) = \sqrt{2}$. The band gap (grey bar) has a relative width of $(\Delta\omega/\omega) = 25.3\%$, centered around frequency $= 0.564$ $[\omega a/2\pi c]$. B) Brillouin zone of the orthorhombic lattice. 8 high symmetry points are indicated, where $\Gamma$ corresponds to (0,0,0), X to $(\frac{1}{2},0,0)$, Y to $(0,\frac{1}{2},0)$, and Z to $(0,0,\frac{1}{2})$. The $\Gamma$ - X and $\Gamma$ - Z directions correspond to the directions parallel to the two sets of pores in the crystal.

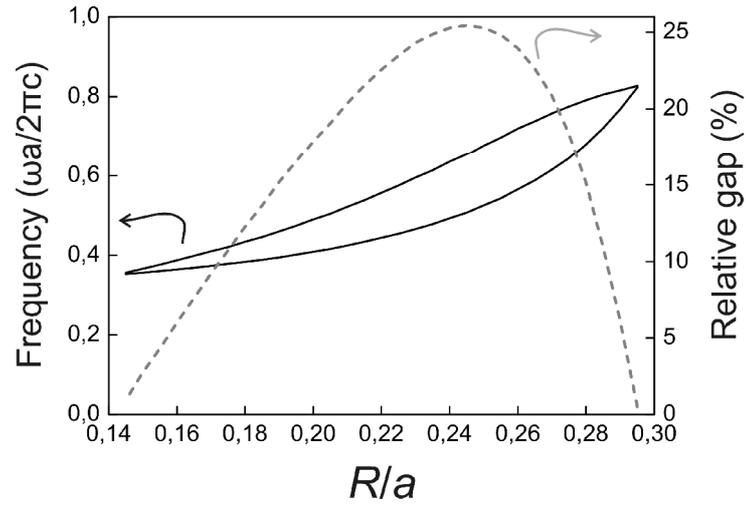

FIG. 3: Calculated relative width of the band gap (right ordinate) and frequencies of band gap edges (left ordinate) versus varying radii of the two sets of pores. $(R/a)$ was varied from 0.14 to 0.30, with 0.005 increments. We find a maximum relative width of the band gap of $(\Delta\omega/\omega) = 25.4\%$ at a central frequency of 0.581 $[\omega a/2\pi c]$ for $(R/a) = 0.245$. For $(R/a)$ values increasingly further away from the optimal value, the relative band gap width is reduced (dashed line). The band gap edges (solid lines) move to larger frequencies with increasing $(R/a)$.

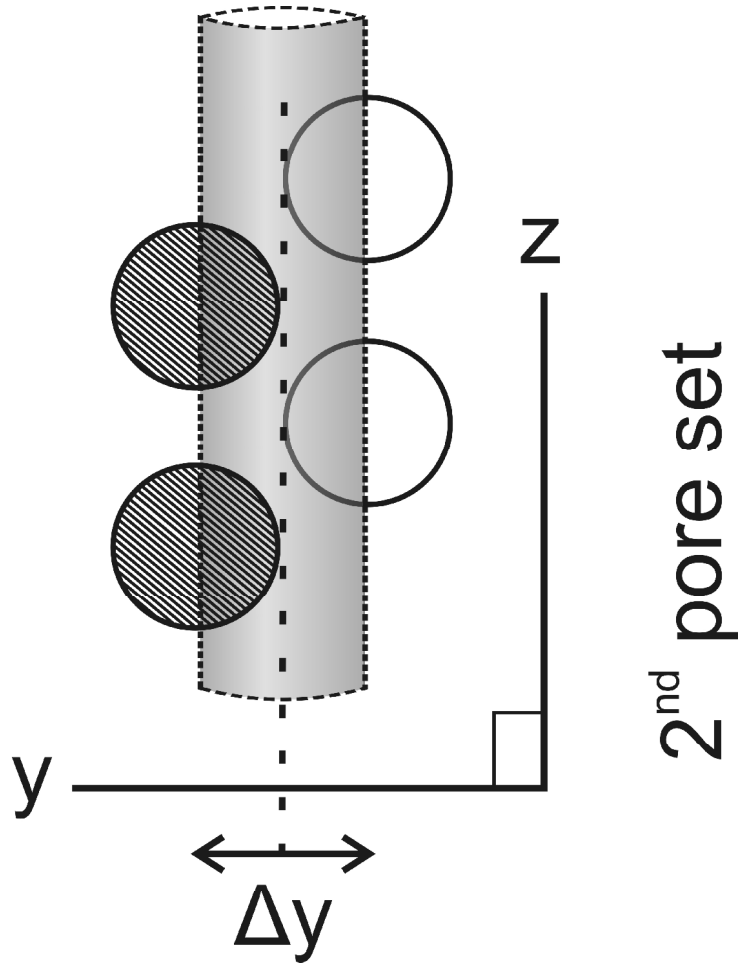

FIG. 4: Representation of the $\Delta y$ displacement that may occur as a result of misalignment of the second set of pores. The first set of pores points "into" the paper and the second set of pores runs parallel to the z-direction. The ideal position for pores in the second set is exactly in the middle of two rows of pores of the first set, indicated by the dashed line. Misalignment of the second set to the left or the right from its ideal position ($\Delta y$) results in reduced relative photonic band gap widths.

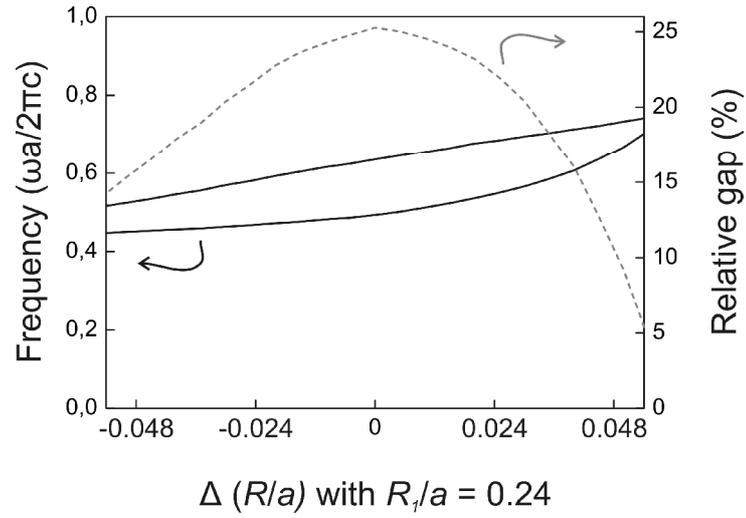

FIG. 5: Calculated relative width of the band gap (dashed line) and band gap edges (solid lines) plotted against $\Delta(R/a)$ for crystals with $(R_1/a) = 0.24$ and $(R_2/a)$ varied from 0.18 to 0.30. The relative width of the band gap decreases as the difference between $(R_2/a)$ and $(R_1/a)$ gets larger. If $\Delta(R/a)$ is changed by $\pm 0.024$, a relative width of the photonic band gap of more than $(\Delta\omega/\omega) = 20\%$ remains. When $\Delta(R/a) = +0.048$, the relative width of the band gap is decreased to less than 30% of the maximum value. For $\Delta(R/a) = -0.048$, around 60% of the maximum relative width of the band gap remains. The band gap edges are lowest at small $\Delta(R/a)$, and increase as $\Delta(R/a)$ increases.

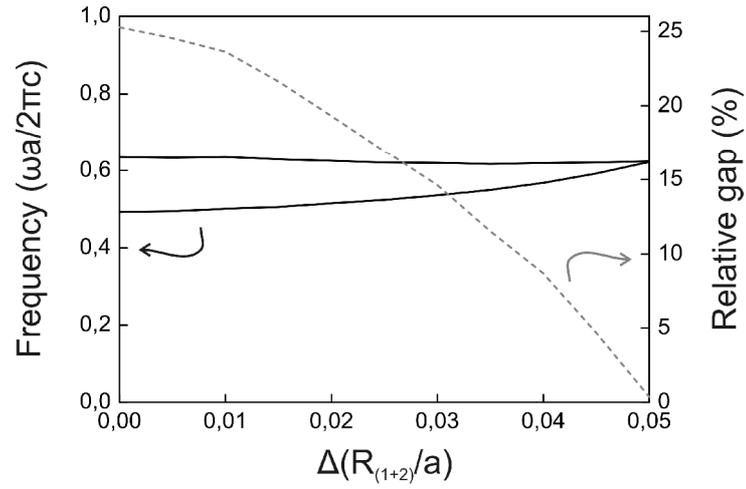

FIG. 6: Calculated relative width of the band gap (dashed line) and band gap edges (solid lines) for crystals with both $(R_1/a)$ and $(R_2/a)$ varying by equal values from $(R/a) = 0.24$. Ratio $(R_1/a)$ is increased, whereas $(R_2/a)$ is decreased. Data is shown for $\Delta(R_{(1+2)}/a)$ up to 0.05, calculated in 0.005 increments. The relative width of the band gap decreases as increases. At a difference of $\Delta(R_{(1+2)}/a) = 0.035$, the relative width of the band gap is halved. At a difference of $\Delta(R_{(1+2)}/a) = 0.05$ the band gap has vanished. The center of the two band gap edges slightly increases as $\Delta(R_{(1+2)}/a)$ increases.

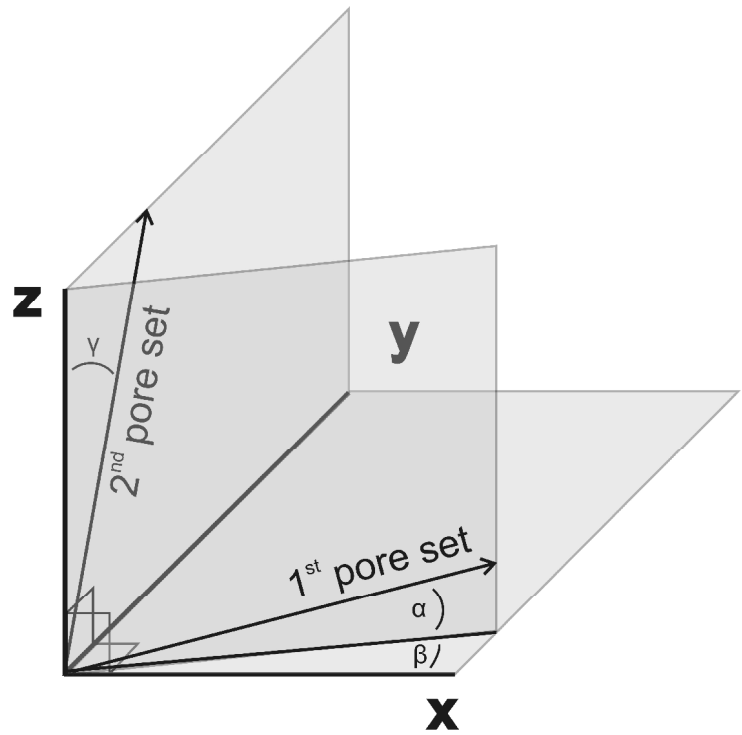

FIG. 7: Schematic of possible angular misalignments that may occur during the fabrication of a three dimensional photonic structure. The x-axis is the ideal direction of the first pore set and the z-axis of the second pore set. Angular misalignments are introduced when the second pores set is etched non- perpendicularly or with a pattern that is rotated with respect to its intended layout.

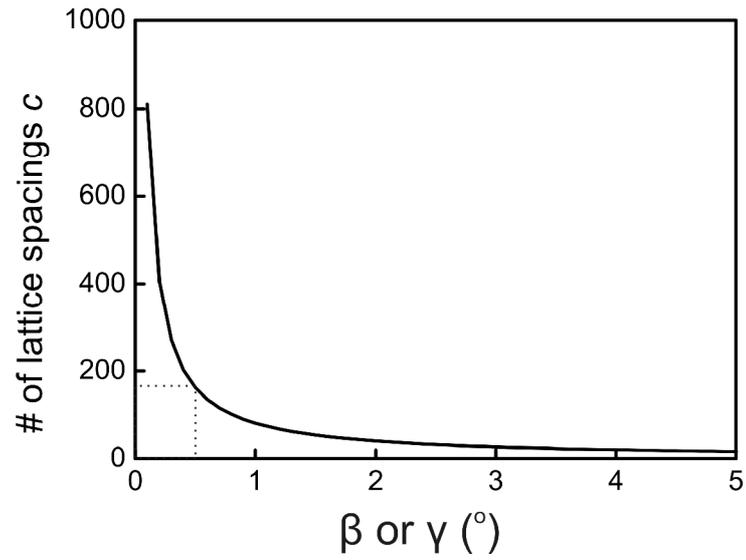

FIG. 8: The number of lattice spacings $c$ that a pore crosses when it traverses one lattice spacing $a$ due to angular misalignment *versus* the misalignment angle $\beta$ or $\gamma$. For $\beta$ or $\gamma = 0.5°$ the structure is periodic over 180 lattice spacings $c$. This value of $0.5°$ will be taken by us as upper limit for these two misalignments individually.

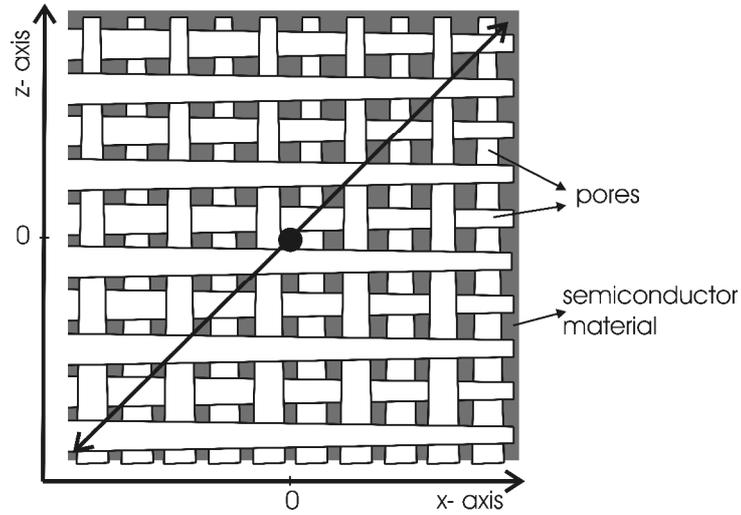

FIG. 9: Schematic cross-section of a structure of 10 pores by 10 pores with pores tapered in both directions. In the third dimension the structure is assumed to be (infinitely) large. In the center of the structure (black dot, [x, z] = [0, 0]) the radii of both pore sets are equal to the desired radius. The diagonal shows positions in the structure were $(R/a)$ is different, while the radii of both pore sets are equal. The arrows indicate the pores and the semiconductor material. In these structures the band gap is chirped and the effect of the tapering on the local gap is studied for positions along the diagonal. The structure shown is approximately 4 times smaller than the one used for the calculations in this section.

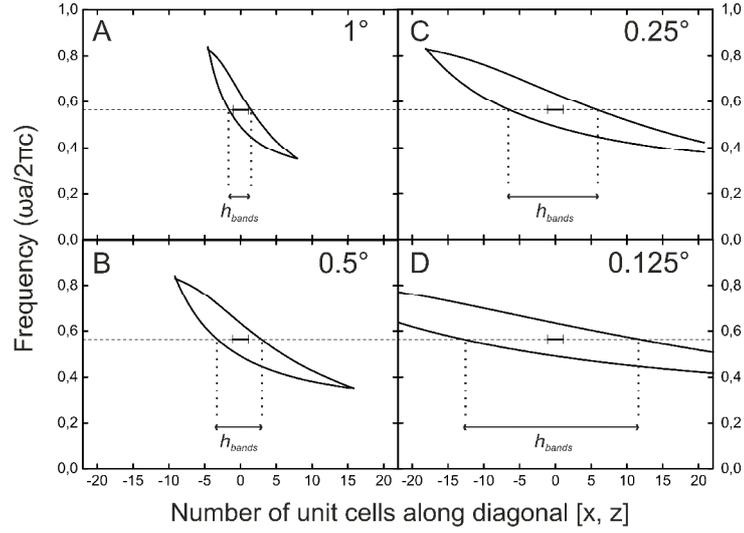

FIG. 10: Estimated edges of the "local photonic band gap" at positions along the diagonal in Figure 9 for inverse woodpiles with pores with different amounts of tapering. The horizontal dashed line displays the relative frequency for which these inverse woodpiles are optimized. The horizontal bar in the center of each plot shows the Bragg length $L_B$. A) Result for 1° tapering. The gap edges are highly curved. Since $h_{bands}$ is almost equal to the Bragg length $L_B$, a band gap can not develop in such structures. B) and C) Results for 0.5° and 0.25° tapering: the gap edges are still highly curved, but a small part of these structures exhibit a band gap. This is indicated by $h_{bands}$, which is larger than $L_B$. D) At a tapering of 0.125°, $h_{bands} = 24.5$ unit cells, which spans more than half the diagonal in a 10 x 10 $\mu m^2$ inverse woodpile with $a = 680$ nm. The estimated transmittance is very low: $T = 1.2 \cdot 10^{-9}$, which is indicative of a strongly photonic structure.

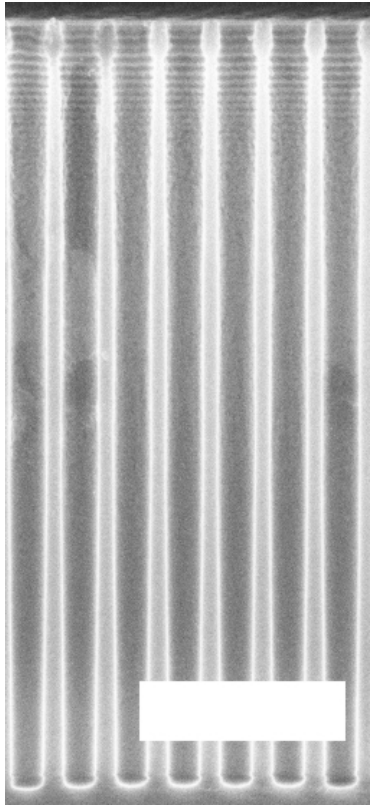

FIG. 11: Cross-sectional view of pores etched using an optimized reactive ion etching process. These pores are in a centered rectangular lattice. They have a depth of $h = 7.5 \pm 0.16$ μm, a diameter of $D_{pore} = 364 \pm 22$ nm, and an interpore distance of 500 nm. The aspect ratio is higher than 20. The tapering of these pores is as low as $0.2 \pm 0.07°$. The scale bar equals 2 μm.

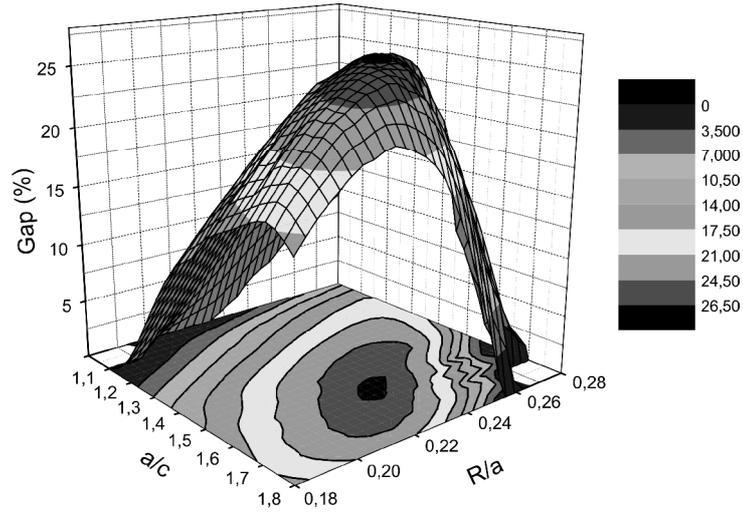

FIG. 12: Interpolated three-dimensional surface- and contour plot of the calculated band gaps versus the ratios $(R/a)$ and $(a/c)$. The Figure shows that there is a large region of $(a/c)$ and $(R/a)$ combinations where the relative width of the band gap is much larger than $(\Delta\omega/\omega) = 26.5\%$, see the black circular area around $(a/c) = 1.6$ and $(R/a) = 0.23$. The broadest photonic band gap of $(\Delta\omega/\omega) = 26.9\%$ is found at $(a/c) = 1.575$ and $(R/a) = 0.23$. The Figure also shows that when a three-dimensional crystal is fabricated from two sets of hexagonal cylindrical pores with high symmetry ( $(a/c) = \sqrt{3}$ ), a much smaller band gap is found.

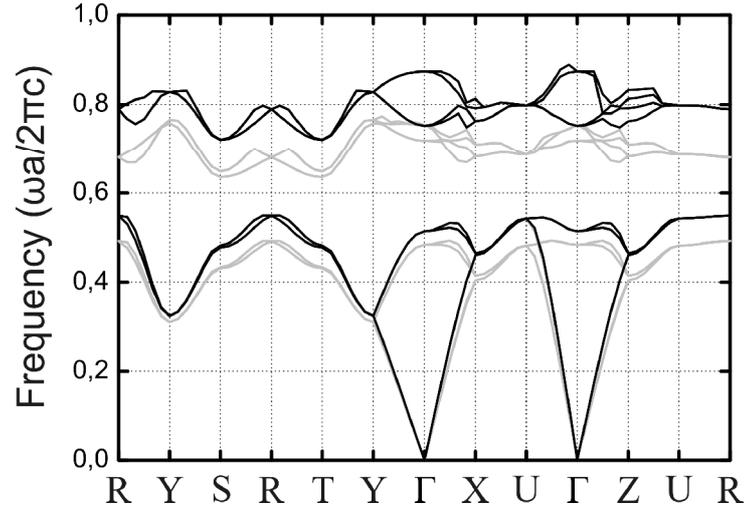

FIG. 13: Calculated band structures of the optimal structure with $(a/c) = 1.575$ and $(R/a) = 0.23$, and the calculated band structures of the crystal with $(a/c) = \sqrt{2}$ and $(R/a) = 0.24$. The central frequency of the optimal structure is $\omega = 0.635\ [\omega a/2\pi c]$, compared to $\omega = 0.564\ [\omega a/2\pi c]$ for the other structure. Apart from the height on the y-axis, the lower bands of the two crystals are equal. For most vectors in the Brillouin zone, the upper bands are also equal, with the exception of the $\Gamma$ - $Y$, $\Gamma$ - $X$, $\Gamma$ - $U$ and $\Gamma$ - $Z$ directions.

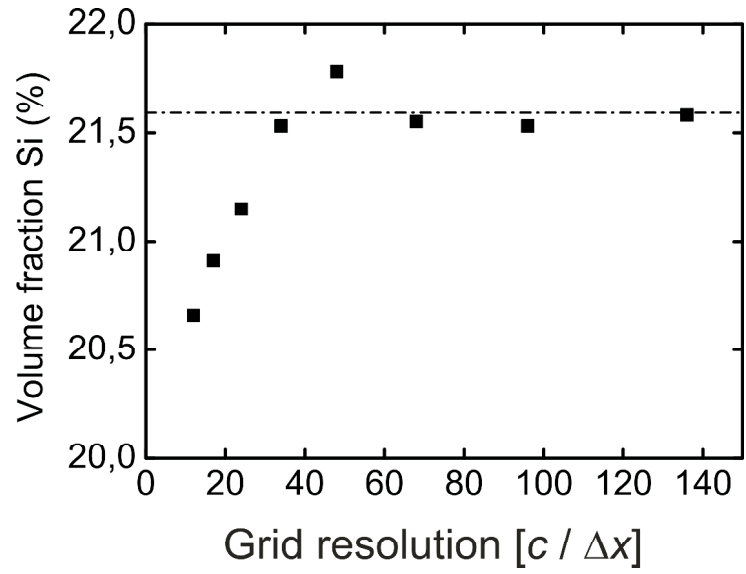

FIG. 14: Calculated volume fraction of silicon versus increasing grid-resolution. The horizontal line indicates the exact volume fraction of silicon in the inverse woodpile structure. At lower grid-resolutions the unit cell is poorly defined, but from a grid-resolution of 68×96×68 and higher, the unit cell is defined with high enough accuracy.

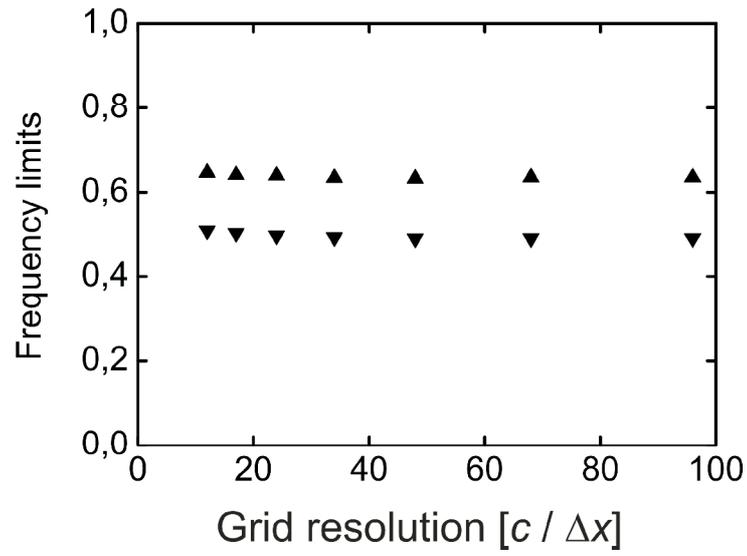

FIG. 15: Calculated frequency limits of the band gap versus increasing grid-resolution. From a grid-resolution of 48x68x48 and higher, the upper- and lower calculated frequency limits are constant.